\documentstyle{aipproc2}

\input epsfc.sty
\def\fdg{\hbox{$.\!\!^\circ$}}

\begin{document}

\title{The SPOrt Project: an Experimental Overview}
\author{S. Cortiglioni$^{\dagger }$, S. Cecchini$^{\dagger }$,
E. Carretti$^{\ddagger }$, M. Orsini$^{\dagger \ddagger }$, R. Fabbri$^{*}$, \\
G. Boella$^{\P }$, G. Sironi$^{\P }$,
J. Monari$^{\S }$, A. Orfei$^{\S }$, 
R. Tascone$^{\S \S}$, U. Pisani$^{\S *}$\\
K.W. Ng$^{\Vert }$, L. Nicastro$^{**}$, L. Popa$^{\dagger \dagger }$,
I.A. Strukov$^{\ddagger \ddagger }$, M.V. Sazhin$^{\P \P }$}

\address{%
$^{\dagger }$I.Te.S.R.E./CNR, Via P. Gobetti 101, I-40129 Bologna, Italy \\
$^{\ddagger }$Dipartimento di Astronomia, Universit\`a di Bologna, Via
Zamboni 33, I-40126 Bologna \\
$^{*}$Dipartimento di Fisica, Universit\`a di Firenze, Via S. Marta 3,
I-50139 Firenze, Italy \\
$^{\P }$Dipartimento di Fisica, Universit\`a di Milano, Via Celoria 16,
I-20133 Milano, Italy \\
$^{\S }$I.R.A./CNR VLBI Radioastronomical Station of Medicina, Via P.
Gobetti 101, I-40129 Bologna, Italy \\
$^{\S \S}$CESPA/CNR c/o Dpt. Elettronica Politecnico di Torino,
c.so Duca degli Abruzzi 24, 10129 Torino, Italy\\
$^{\S *}$Dpt. Elettronica Politecnico di Torino,
c.so Duca degli Abruzzi 24, 10129 Torino, Italy\\
$^{\Vert }$Institute of Physics, Academia Sinica, Taipei, Taiwan 11529,
R.O.C. \\
$^{**}$I.F.C.A.I./CNR, Via U. La Malfa 153, I-90146 Palermo, Italy \\
$^{\dagger \dagger }$Institute of Space Sciences, R-76900
Bucharest-Magurele, Romania \\
$^{\ddagger \ddagger }$Space Research Institute (IKI), Profsojuznaja ul.
84/32, Moscow 117810, Russia \\
$^{\P \P }$Schternberg Astronomical Institute, Moscow State University,
Moscow 119899, Russia}

\maketitle

\begin{abstract}
The Sky Polarization Observatory (SPOrt) is presented as a project aimed
to measure the diffuse sky polarized emission, from the International
Space Station, in the frequency range
20--90 GHz with $7^\circ$ of $HPBW$. The SPOrt experimental configuration is
described with emphasis on the aspects that make SPOrt the first
European scientific payload operating at microwave wavelengths.
\end{abstract}

\section*{Introduction}

The first idea of the SPOrt project was born in 1996 in reply to the ESA
Call for Ideas for the Early Utilization Phase of the International Space
Station (ISS)(1). At that time SPOrt was aimed to measure
only the polarization of the Galactic Background in the frequency range
10--30 GHz, in order to fill the gap in observational data in a region of
particular interest for Cosmic Background Radiation (CBR) measurements. 
For many reasons (see for example Cortiglioni et al. (2) and Fabbri et 
al. (3)) the current SPOrt configuration is quite
different from the original one, including among its objectives also 
the attempt to measure CBR
linear polarization.  This more ambitious scientific goal can be reached
only with an improved design, which is characterized by:

\begin{itemize}
\item 20--90 GHz frequency coverage, in order to optimize foreground
 subtraction;

\item the same beamwidth for all channels ($HPBW=7^\circ$): no normalization is
 needed in data analysis;

\item Phase Switched Correlation Radio-Polarimeters (PSCRP) with direct
 amplification chains, which provide simultaneously $Q$ and $U$ Stokes
 parameters almost free from offsets, drifts and $1/f$ noise;

\item very simple optical design without any reflector optics in order to have
 rejection to unpolarized components as high as possible;

\item the best available low noise front end amplifiers to improve
 instantaneous sensitivity;

\item observing conditions as stable as possible in order to allow long time
 integrations ($10^4 \div 10^5$ s);

\item sky coverage $>80$\%.
\end{itemize}

Other experiments are currently attempting to measure CBR polarization from
ground (see for example Sironi et al., 1998 (4) and Keating et al., 1998 (5))
but with limited sky and frequency coverage. Also the measurements technique
is quite different from that adopted in SPOrt, where dedicated components
are used to optimize the experimental configuration.
Most challenging aspects of the SPOrt project, apart from the extremely
short realization time imposed by the ISS schedule, can be summarized by: 

\begin{itemize}
\item realization of an antenna system suitable to separate the two circular
components of the observed radiation with low ($<-45$ dB)
cross-polarization;

\item realization of a radio-polarimetric chain able both to improve the total
unpolarized component rejection up to $-70$ dB and to reduce as much as
possible the total contribution of systematics (drift, offsets, $1/f$ noise);

\item realization of an overall experiment structure suitable to fit the ISS
constraints (mechanical, thermal, observational).
\end{itemize}

\section*{The SPOrt experiment}

Following the above ideas the SPOrt design is based on four separate
polarimeters at 22, 32, 60 and 90 GHz characterized by: 
\begin{enumerate}
\item Corrugated Feed Horn Antennas (CFHA) with $HPBW=7^\circ$, extremely 
low sidelobes and cross-polarization;
\item Iris Polarizer (IP) followed by an Orthomode Transducer (OMT), 
to provide separate Left Hand Circular (LHC) and Right Hand Circular (RHC) 
components;
\item two low noise Phase Switched Radiometric Chains (PSRC);
\item Analog Correlation Unit (ACU);
\item Phase Synchronous Detection Unit (PSDU);
\item Post Detection Unit (PDU).
\end{enumerate}
 
The signal collected by the antennas is divided into its two circularly
polarized components, LHC and RHC, by both the IP and the OMT. 
The OMT provides then two outputs,
which are independently amplified by two PSRCs and then correlated by the
ACU in order to get outputs proportional to both $Q$ and $U$ Stokes
parameters. In this way it is possible to have an instantaneous full
characterization of the linear polarization of the observed sky. 

\begin{figure}
\centerline{%
\epsfxsize=\textwidth  
\epsfbox[76 221 441 753]{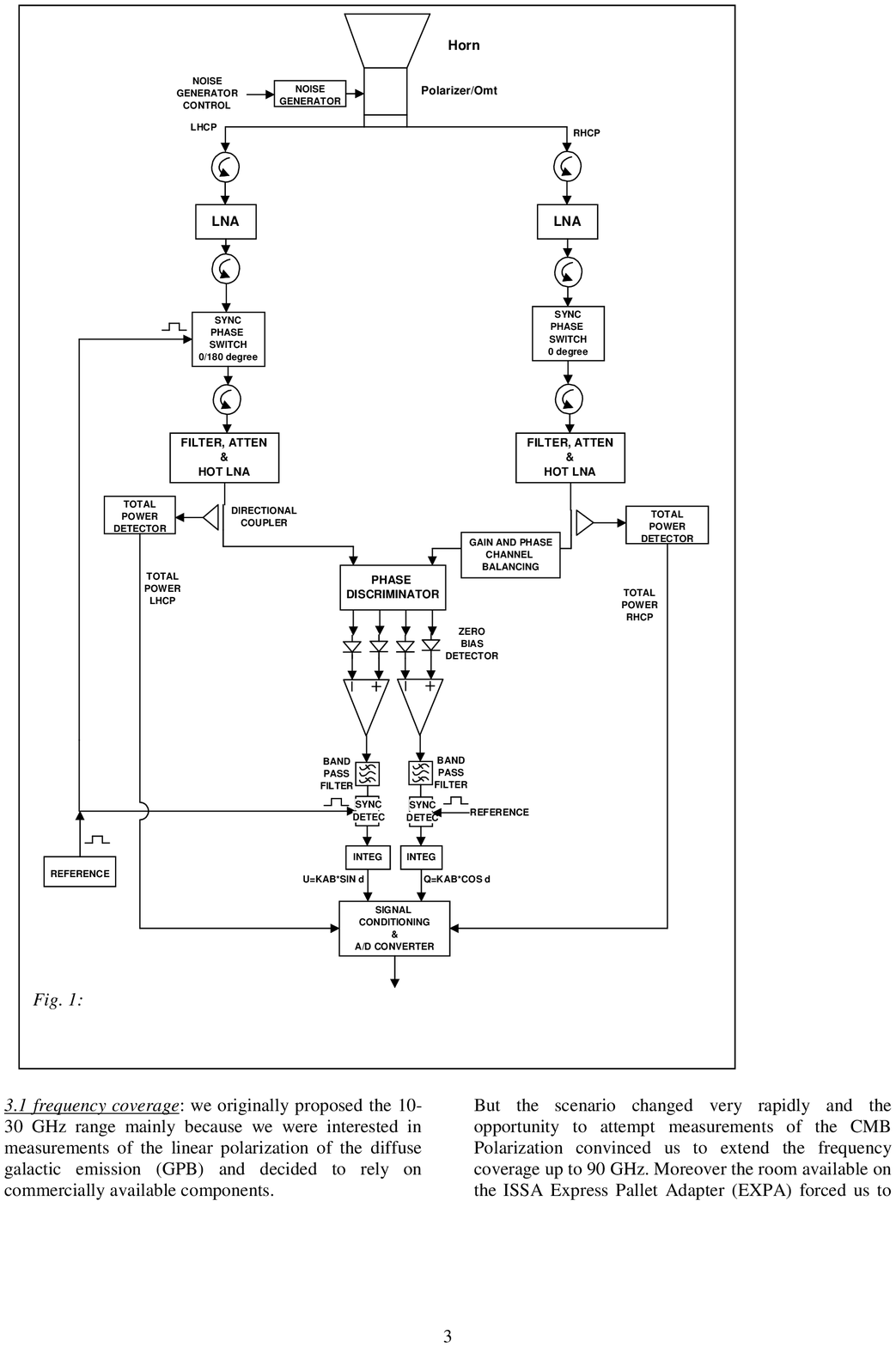} }
\caption[]{Schematic block diagram of the SPOrt radiometers.}
\label{block}
\end{figure}

From a bandwidth point of view, a good sensitivity calls for a large
absolute value but relatively small with respect to the central frequency
to avoid degradation of the component performance versus the bandwidth. 
For these reasons the SPOrt radiometers use RF bandwidths of 10\%,
avoiding a down conversion and correlating the signals at RF level.
The capability to reach the long term sensitivity at
$\mu$K level relies on two main characteristics of the radiometers: i) 
noise temperature as low as possible as well as non gaussian noise at
$\mu$K level (after off-line integration); ii) the best characterization 
of performances so that all systematic effects can be removed. The
correlation technique, in fact, should provide $Q$ and $U$ Stokes parameters
free, in principle, from the $1/f$ noise coming from the low noise amplifiers as
well as from uncorrelated gain variations of the two receiver channels. 
Moreover the modulation of the RF signal helps to remove
undesired effects such as output drifts, flicker noise coming from post
detection electronic section and spurious cross-polarization from non
ideal devices.  Taking into account all these aspects a schematic block diagram
of each SPOrt channel is shown in Figure \ref{block}. It could be divided into
three principal parts: the antenna system, the low noise 
PSRC, the correlation-detection-demodulation section (ACU + PSDU).

\subsection*{Antenna system}

\begin{figure}[ht]
\centerline{
\epsfxsize=10cm
\epsfbox[136 142 456 401]{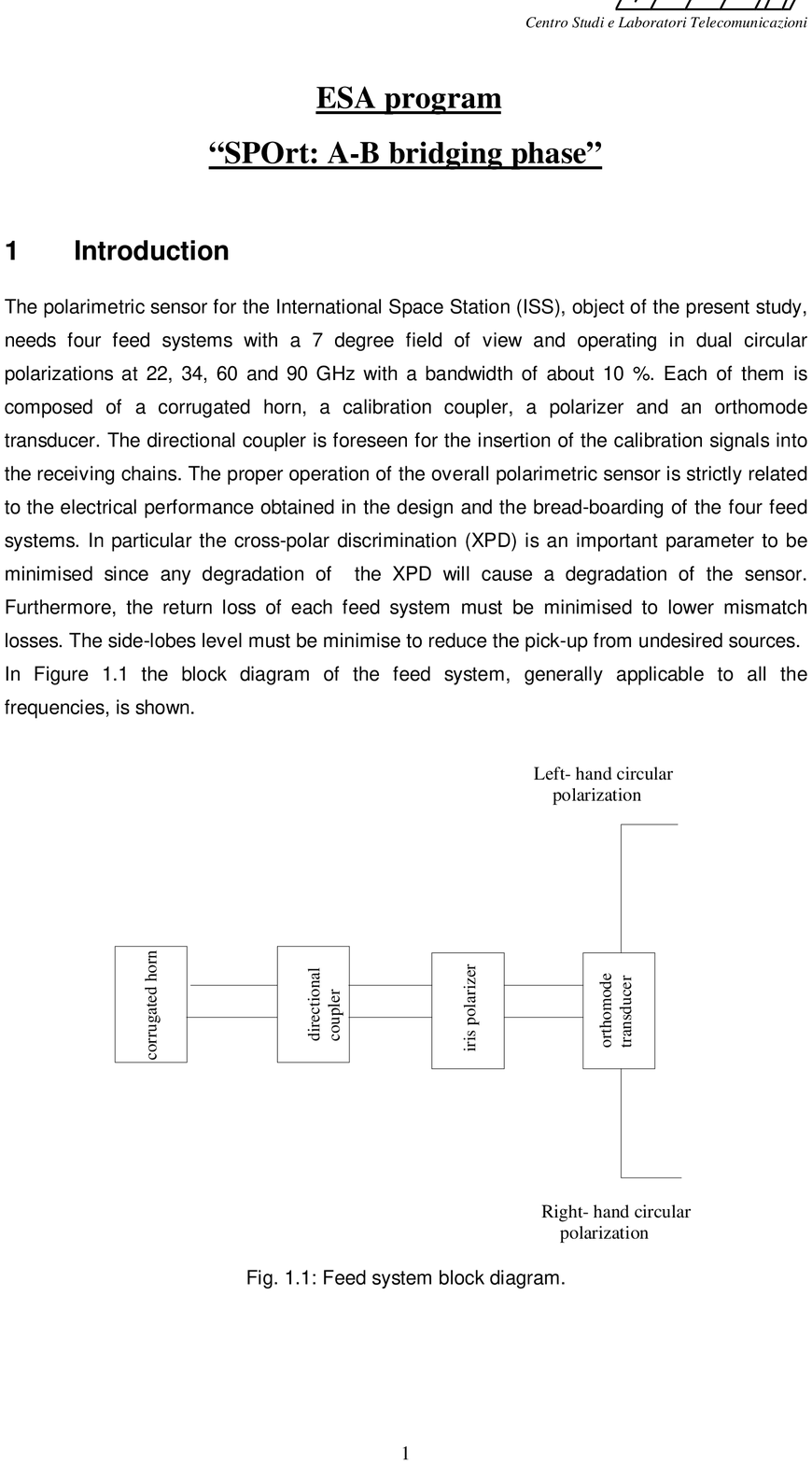} }
\caption[]{SPOrt antenna system.}
\label{asys}
\end{figure}

The antenna system will include: CFHA, IP, OMT and a coupling device (e.g. a 
waveguide with directional coupler) to inject a known noise calibration 
signal (see Figure \ref{asys}).
All these parts must be included in a unique design to optimize overall
characteristics.

\begin{figure}
\centerline{
\epsfxsize=10cm
\epsfbox{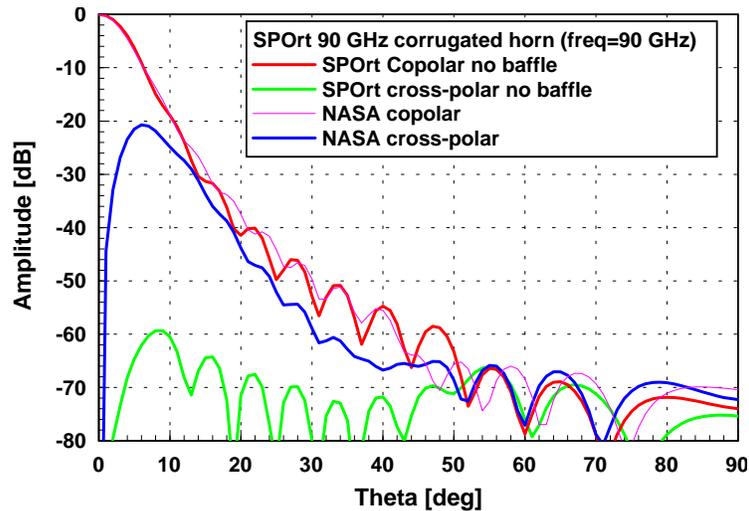} }
\caption{SPOrt antenna pattern compared to COBE/DMR.}
\label{aptt}
\end{figure}

The most relevant SPOrt antenna system specifications 
at all frequencies can be summarized in: a) 10\% 
bandwidth, two circular polarizations and 
$7^\circ$ $HPBW$ beamwidth, b) an insertion loss $< 0.5$ dB, c) a return loss 
$< -25$ dB, d) a sidelobe level $< -60$ dB and e) a cross polarization 
level $< -45$ dB. Figure \ref{aptt} shows an example of the current status
of the antenna system design for SPOrt. Both the copolar and the cross-polar
pattern of the 90 GHz feed of COBE/DMR are also shown for comparison.

%

\subsection*{Low Noise PSRC}

Low Noise Amplifiers (LNA) represent the most critical component, 
together with the insertion loss of the antenna system, 
to obtain instantaneous sensitivities $< 2$ mK s$^{\frac{1}{2}}$ 
as required to meet SPOrt scientific goals (3).
Two LNA
stages are needed (for a total $\sim$ 70 dB gain): the first one 
will be cooled
at 80 K and the gain must be enough to mask the noise of the following
components; the second amplifier, together with the rest of the receiver,
is included in the warm unit ($\sim$ 300 K).
The ($0^\circ-180^\circ$) phase switch is part, together with the 
PSDU, of the lock-in chain, which has the task to minimize gain 
drifts, offsets, $1/f$ noise generated inside the loop. This method will 
allow also to remove the cross-polarization generated in the HPD as well 
as part of the spurious effect due to the unpolarized component (see 
sections below). 

\subsection*{The polarimetric unit}

The polarimetric unit is essentially the ACU and includes
 (see Figure \ref{block}):

\begin{itemize}
\item Hybrid Phase Discriminator (HPD);
\item square law detectors;
\item differential amplifiers plus integrator.
\end{itemize}
\begin{figure}[ht]
\centerline{
\epsfxsize=8cm
\epsfbox{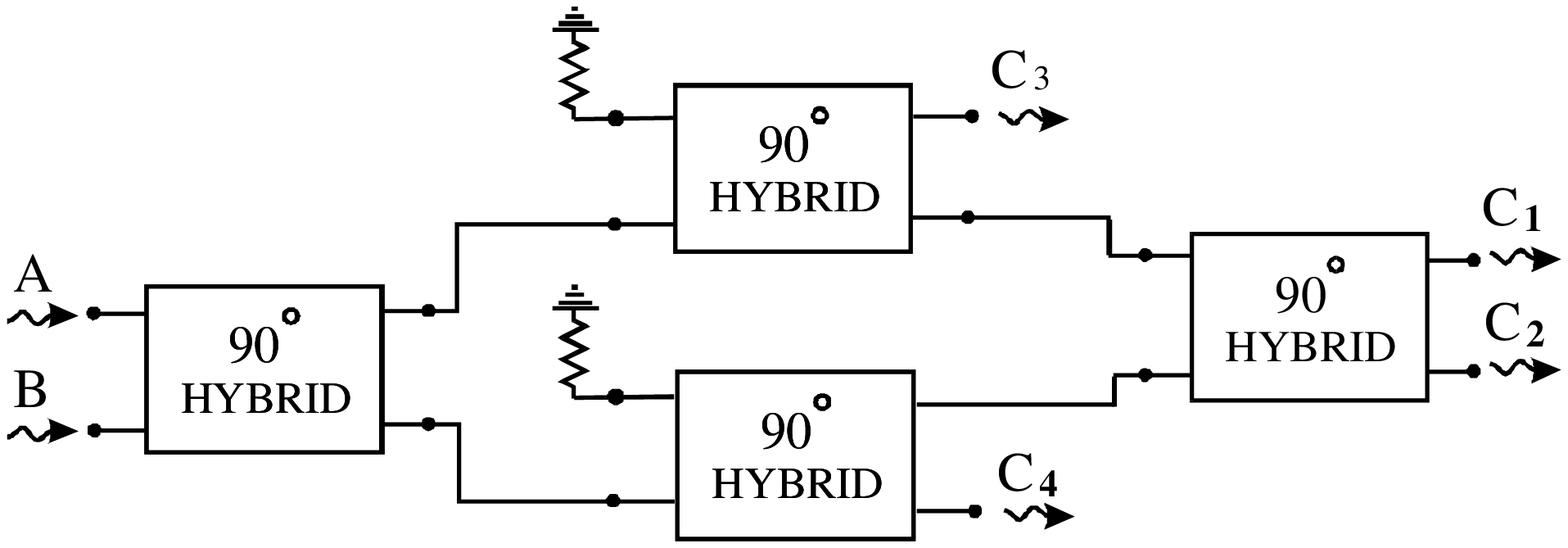} }
\caption{Equivalent circuit of SPOrt HPDs.}
\label{hpd}
\end{figure}
The HPD is a passive microwave circuit consisting of four Hybrid 
directional couplers and one $90^{\circ}$ differential phase shifter, 
that processes the signal in order to have four outputs.  
The HPD equivalent circuit is shown in Figure \ref{hpd}.
After square law detection the four outputs: 
$$
\begin{array}{lclcl}
V_{1} & = &
\left|c_{1}\right|^{2} & = &
\frac{1}{4} \left[
\left|A\right|^{2} + \left|B\right|^{2}
- 2 \Re \left\{ A B^{*} \right\}
\right] \\ \\
V_{2} & = &
\left|c_{2}\right|^{2} & = &
\frac{1}{4} \left[
\left|A\right|^{2} + \left|B\right|^{2}
+ 2 \Re \left\{ A B^{*} \right\}
\right] \\ \\
V_{3} & = &
\left|c_{3}\right|^{2} & = &
\frac{1}{4} \left[
\left|A\right|^{2} + \left|B\right|^{2}
+ 2 \Im \left\{ A B^{*} \right\}
\right] \\ \\
V_{4} & = &
\left|c_{4}\right|^{2} & = &
\frac{1}{4} \left[
\left|A\right|^{2} + \left|B\right|^{2}
- 2 \Im \left\{ A B^{*} \right\}
\right]
\end{array}
$$
%
%
%
%
(where $A$ and $B$ represent previously called LHC and RHC after
amplification) are then properly differentiated to get two outputs

\begin{eqnarray}
\Re \left\{ A B^{*} \right\} & \equiv & |AB| \cos \phi \\
\Im \left\{ A B^{*} \right\} & \equiv & |AB| \sin \phi
\end{eqnarray}
%
%
which after integration will give time averaged values
directly related to $Q$ and $U$ Stokes parameters.  

The HPDs working at the four SPOrt
frequencies will be especially designed and manufactured with waveguide
technology for the SPOrt experiment because they are not commercially available.
To optimize performances, an integrated waveguide 
solution will be adopted. In particular, the coupling between the rectangular 
waveguides of the directional couplers and the phase delay of the phase
shifter will
be obtained by using a perforated septum placed on the E-plane. In this way i) 
the coupling between waveguides will be in the H-plane and ii) the waveguides 
are 
5-folded in the E-plane to obtain a compact solution which has been called 
``5-floor-HPD". The whole configuration was studied to be manufactured by 
electroerosion technology so that stringent tolerances can be maintained.

At the four frequency bands, both hybrid directional couplers and phase 
shifters, must have a return loss of 40 dB, 
an isolation of 40 dB, the balance function ratio less than 0.1 dB 
and a phase error of $\pm$ 0.1 deg.  
The specifications will be maintained in the 10\% bandwidth. 
Inside each device will be also integrated a filter to properly 
shape the input band and a directional coupler to split part of the 
incoming power for the total power detection (see Figure \ref{block}).
One of the most important specifications
for the HPD is the rejection of the unpolarized signal 
coming from the sky, such as that due to the CBR ($\sim$ 3K).
This unwanted contribution is eliminated by the difference 
carried out by two differential amplifiers. For this reason, 
it is very important to maintain a high level of symmetry in the 
whole device. It can be shown that the cross signal
generated from the devices inside the modulation loop 
(see Figure \ref{block}) is thrown away because
of the modulation technique. This technique is also adopted to minimize
both characteristics dispersion and $1/f$ noise coming from other components
(i.e. diodes and differential amplifiers, etc ...). 

\section*{The problem of the spurious polarization}
\label{sec:crosspol}
%

The output expressions in the previous section are only an idealization.
In the real
case, due to cross-polarization effects, part of the unpolarized incoming
radiation will appear as polarized. In addition, the two
Stokes parameters $Q$ and $U$ will pollute each other so that a practical
expression of the correlated outputs should be: 
\begin{eqnarray}
Q_t & = & Q + a\, U + b\, |A|^2 + b\, |B|^2  \label{qt1}\\
U_t & = & U + c\, Q + d\, |A|^2 + d\, |B|^2  \label{ut1} 
\end{eqnarray}
where $a$, $b$, $c$ and $d$ are coefficients to be measured during antenna 
characterizations. The terms $|A|^2$ and $|B|^2$ are due to the
cross-polarization generated in the antenna system. 
After PSDU, only cross-contributions from the antenna system will
remain, and very much effort is currently 
devoted to minimize and characterize such residuals.
Moreover the coefficients in the
equations above are $\ll 1$ so the term $a\, U$ in equation (\ref{qt1}) and
the term $c\, Q$ in
equation (\ref{ut1}) are negligible with respect to $Q$ and $U$ respectively.
Thus it can be written:
\begin{eqnarray}
Q_t & = & Q + b\, |A|^2 + b\, |B|^2 \label{qt2}\\
U_t & = & U + d\, |A|^2 + d\, |B|^2 \label{ut2}
\end{eqnarray}
The coefficients $b$, $d$ represent the cross-polarization of the antenna
system.

It can be shown that the coefficients $b$, $d$ are due 
to non-orthogonality between LHC and RHC. In other 
words two components elliptically polarized, but rigorously 
orthogonal, do not generate spurious correlated signals. 
In particular, as far as the spurious correlated signals are 
concerned,  the parameter that must minimized is not the 
cross-polarization level of the antenna system but rather the  
following expression:
$$
\frac{4}{\pi \Delta \theta^2_{-3dB} \Delta \omega} \int_{\Delta \omega}
\int_0^{\pi} \int_0^{2 \pi} \underline{h}_A \left( \omega, \theta, \varphi 
\right) \cdot \underline{h}_B^{*} \left( \omega, \theta, \varphi
\right) sin \theta d \varphi d \theta d \omega 
$$
where $\Delta \theta_{-3dB}$ is the HPBW of the antenna, $\Delta \omega$
is the bandwidth and $\underline{h}_A$ ($\underline{h}_B$) is the vector
radiation pattern (it has components in $\theta$ and $\varphi$ directions)
corresponding to the port A (B) of the Orthomode Transducer, normalized 
to its maximum value.  As a comment of the above expression, one can 
observe that the scalar product between the two vectors $\underline{h}_A$
and $\underline{h}_B$ which describe the radiation field, is weighted 
by the radiation pattern through the integration on the ($\theta - \varphi$)
directions and cancellation effects may occur. Thus, the optimization 
of the instrument design with respect to the orthogonality between 
the two polarizations of the antenna allows to reduce the level of the
spurious correlated signal.
Usual values for $b$, $d$ will range around 0.001, so an offset of
few mK will appear at the output for each on board integration time.
An accurate characterization and stability of this systematic effect 
is needed to allow final sensitivities at $\mu$K level. 


\section*{The expected pixel sensitivity}

The orbit of the ISS was calculated (with some assumptions
on the flight parameters not yet specified by NASA) to simulate
the path of the SPOrt radiometers across the sky and to compute
total integration time versus sky position.
This was done assuming the radiometers pointing at the zenith of the 
ISS Express Pallet even if the developed code can take into account
any inclination with respect to this direction.

The orbit has a quasi-sinusoidal shape in the RA--DEC plane and its
amplitude is equal to the inclination of the ISS orbit, 51\fdg6. 
In fact the precession effect translates into an RA shift of 
$360^\circ$/(Prec. period)/(Orbits per day) = 0\fdg32 per
orbit. Such an orbit has the following characteristics: the same 
point is scanned $\sim 2 \times 7^\circ / 0\fdg32 = 45$ times 
per precession period; the sky portion
scanned by SPOrt in half and all the precession period 
are $\sim 70$\% and $\sim 82$\% respectively. 

\begin{figure}[ht]
\centerline{
\epsfxsize=8cm
\epsfbox{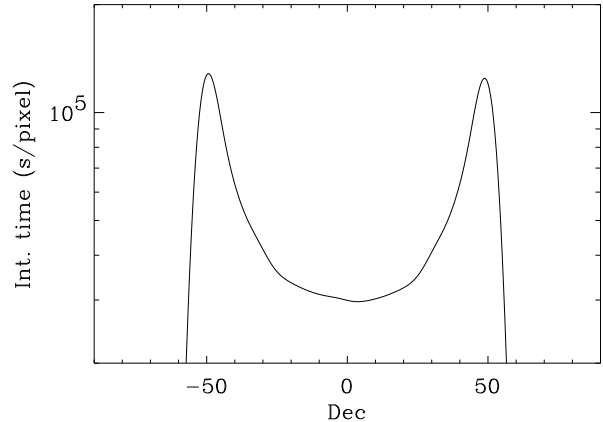} }
\caption{Integration time across the sky vs declination.}
\label{intt}
\end{figure}

The expected integration time is shown in Figure \ref{intt} for
the whole observed declination interval. 
The map shows symmetry
along the $\delta = 0$ axis and peaks at $\pm 51\fdg6$, the values of
integration time do not depend on the right ascension coordinate.
Highest integration times/pixel are far from the celestial equator 
because of:
\begin{enumerate}
\item a slower motion in the declination axis approaching the $\pm 51\fdg6$
\item the $1/\cos \delta$ beam deformation: meridians are closer each other 
 at high declinations and the beam width seems larger, so the number of 
 independent pixels is lower.
\end{enumerate}

The sensitivity for each pixel was then calculated assuming that the 
radiometer equation
$$
\Delta T_{\rm rms} = \frac{\sqrt{2} \cdot T_{\rm sys}}{\sqrt{\Delta \nu \cdot 
\tau}}
$$
holds for the expected integration time, where $\Delta \nu$ is the bandwidth
(10\% of $\nu$), $\Delta \tau$ is the integration time, 
$T_{\rm sys}$ is the system temperature and the factor $\sqrt{2}$ 
comes from the radiometer's configuration (correlation radiometer).
Due to various effects, such as stray-light from solar panels, Sun and Moon
entering into the sidelobes, shuttle 
operations and so on, it would not be possible to use the full SPOrt data set.
As a conservative estimate we consider a 50\% observing efficiency.
Basic data on the expected final sensitivity for the single pixel and
for the full sky are reported in Table \ref{performance}; $T_{\rm sys}$ are 
computed from simulations based on preliminary SPOrt specifications and 
available components. Current development activities would indicate that 
some margin of improvement is still left.
\begin{table}
\caption[ ]{SPOrt expected sensitivity after 1.5 yr lifetime.} 
\label{performance}
\begin{tabular}{lddddd}
{\boldmath $ \nu$ } & \multicolumn{1}{c}{\boldmath $ T_{\mathrm{sys}}$}
& $\Delta T_{\rm rms}$
& \multicolumn{2}{c}{%
{\boldmath
$P_{\mathrm{pix}}$}
($\mu $  K)\protect{\tablenote%
{Computed for 50\% efficiency and 10\% frequency bandwidth.
}}}%
& \multicolumn{1}{r}{%
{\boldmath $P_{\mathrm{rms}}$} {\bf (FS)}\protect{\tablenote%
{Full sky (FS) coverage is 81.7\% of 4$\pi $ sr, including 662
pixels.}}}
\\  \cline{4-5}
(GHz) & \multicolumn{1}{c}{(K)} & \multicolumn{1}{c}{(mK s$^{\frac{1}{2}}$)} &
 \multicolumn{1}{c}{min}
 & \multicolumn{1}{c}{max}
& \multicolumn{1}{c}{($\mu $K)} \\
\tableline
22 & 66. & 2.0 & 7.85  & 16.3 & 0.52  \\
32 & 82. & 2.0 & 8.09  & 16.8 & 0.54  \\
60 & 115. & 2.0 & 8.57  & 17.8 & 0.57  \\
90 & 167. & 2.0 & 9.82  & 20.4 & 0.65
\end{tabular}
\end{table}

%

\section*{Conclusions}

The SPOrt project, after its selection by ESA (end 1997), is currently
under phase B. Part of the 1999 activity will be devoted to build
breadboards at 22 and 90 GHz in order to make engeneering tests before
to proceed towards the Flight Model realization
(payload delivery to ESA/NASA is by the end of year 2000). The very tight ISS
schedule have forced the SPOrt team (Scientific collaboration plus
Industrial consortium) to adopt a new philosophy for the project
realization, which may be resumed as ``faster, better and
cheaper". Since SPOrt represents the first European scientific payload
at microwave frequencies, it will be also a unique opportunity to test
technological solutions that could be the base for future and more
ambitious CBR space project. SPOrt, in fact, shall begin to operate 
just after the MAP satellite (2001) and before PLANCK (2007), so 
playing an important bridging role between NASA and ESA CBR space 
programs. PLANCK-LFI may also benefit of some technological solution, that 
can be tested in-flight by using SPOrt opportunities as well as of 
scientific results especially related to galactic foregrounds. SPOrt would play 
also an important role to promote technological developments that can be
used for ground application in Radioastronomy at short times.
\acknowledgments
This work, as well as the SPOrt project,
is supported by Agenzia Spaziale Italiana (ASI). The European Space Agency
has supported SPOrt's A--B bridging phase under EPI industrial contracts.
M.V.S. thanks the CentroVolta--LandauNetwork for financial support. I.A.S.
thanks the Astronomy Dept. of the Bologna University and the ITeSRE/CNR
for the support given to his participation to SPOrt activities.

\end{document}